\documentstyle[preprint,aps]{revtex}
\begin{document}
\draft
\tightenlines
\title{Critical-point scaling function for the specific heat of a Ginzburg-Landau superconductor}
\author{Dominic J. Lee\thanks{Present address: Department of Physics, Simon Fraser University, Burnaby, BC, Canada} and Ian D. Lawrie}
\address{Department of Physics and Astronomy, University of Leeds, Leeds LS2 9JT, England}
\maketitle
\begin{abstract}
If the zero-field transition in high temperature superconductors such as $\textrm{YBa}_2\textrm{Cu}_3\textrm{O}_{7-\delta}$ is a critical point in the
universality class of the 3-dimensional XY model, then the general theory of critical phenomena predicts the existence of a critical region in which
thermodynamic functions have a characteristic scaling form.  We report the first attempt to calculate the universal scaling function associated
with the specific heat, for which experimental data have become available in recent years.  Scaling behaviour is extracted from a renormalization-group
analysis, and the $1/N$ expansion is adopted as a means of approximation.  The estimated scaling function is qualitatively similar to that observed
experimentally, and also to the lowest-Landau-level scaling function used by some authors to provide an alternative interpretation of the same data.
Unfortunately, the $1/N$ expansion is not sufficiently reliable at small values of $N$ for a quantitative fit to be feasible.
\end{abstract}
\pacs{64.60.Fr, 74.25.Bt}
\section{Introduction}

In recent years, a considerable body of experimental evidence has accumulated to suggest that the zero-field transition in certain
high-temperature superconductors, most notably $\textrm{YBa}_2\textrm{Cu}_3\textrm{O}_{7-\delta}$ (YBCO), is a critical point in the universality
class of the 3-dimensional XY model \cite{inderhees,salamon,schneider,overend,overend2,cooper,moloni}.  If this is the case, then, in the presence of a
sufficiently small magnetic field $B$, the specific heat is expected to have a singular part which exhibits the scaling behaviour
\begin{equation}
C_{\mathrm{sing}}(T,B)=B^{-\alpha/2\nu}{\cal C}(x)\,,
\label{shscaling}
\end{equation}
where $\alpha\approx-0.013$ and $\nu\approx 0.67$ are critical exponents and the scaling variable is $x=(T-T_{\mathrm{c}})B^{-1/2\nu}$.  Similar scaling
forms are expected for other thermodynamic quantities, such as the magnetization.  In the limit $B\to 0$, the scaling function must behave as
${\cal C}(x)\to C_{\pm}\vert x\vert^{-\alpha}$, so that  $C_{\mathrm{sing}}(T,0)=C_{\pm}\vert T-T_{\mathrm{c}}\vert^{-\alpha}$, where $+$ refers to
$T>T_{\mathrm{c}}$ and $-$ to $T<T_{\mathrm{c}}$. For YBCO, zero-field measurements of the specific heat presented by several authors seem to agree
well with this prediction \cite{overend,junod} and to be consistent with the universal values of $\alpha$ and of the amplitude ratio $C_+/C_-$ as determined
by precision measurements of the superfluid transition in $^4$He, which is also in the universality class of the 3-dimensional XY model
\cite{lipa1,lipa2}. A claim has recently been made that the zero field singularity is actually characterized by different exponents $\alpha_+$ and $\alpha_-$,
above and below $T_{\mathrm{c}}$, which would not be consistent with any ordinary type of critical point \cite{charalambous}.  It has also been argued,
though, that this conclusion rests on an inappropriate background subtraction \cite{ramallo}.

In a nonzero applied field, one can test the scaling form (\ref{shscaling}) by the extent to which data for $B^{\alpha/2\nu}C_{\mathrm{sing}}(T,B)$
collapse to a common curve when plotted as a function of $x$. Here, matters are complicated by the fact that a different kind of scaling behaviour
\begin{equation}
C(T,B)\approx{\cal C}_{\mathrm{LLL}}(x_{\mathrm{LLL}})
\label{lllscaling}
\end{equation}
is expected when only the lowest Landau level is significantly occupied \cite{bray,thouless}. Here, the scaling variable is $x_{\mathrm{LLL}}=
(T-T_{\mathrm{c}2}(B))/(TB)^{2/3}$, where $T-T_{\mathrm{c}2}(B)$ or, equivalently, $B=B_{\mathrm{c}2}(T)$ is the upper critical field of the Ginzburg-Landau
theory. Since $\alpha$ in (\ref{shscaling}) is very small, and $1/2\nu\approx0.75$, the two predictions are rather hard to distinguish.  Some authors claim
that lowest-Landau-level scaling works just as well as, or indeed better than, critical point scaling
\cite{welp,pierson1,pierson2,pierson3,jeandupeux,roulin,junod2}.  For ${\rm HgBa}_2{\rm Ca}_2{\rm Cu}_3{\rm O}_{8+\delta}$ \cite{carrington}, specific heat
data appear to collapse to a common curve  when plotted in the form of (\ref{shscaling}), but the scaling function ${\cal C}(x)$, which ought to be universal, is
apparently rather different from that found for YBCO. For ${\rm LuBa}_2{\rm Cu}_3{\rm O}_7$, the authors of Ref.\cite{nanda} find that a {\it two-dimensional}
lowest-Landau-level scaling form best fits the data, though they claim that it is also consistent with 3-dimensional XY scaling for fields below 1T. Most recently,
Junod {\it et al} \cite{junod3} have concluded that optimally-doped YBCO is the only material to show convincing evidence of critical-point scaling.

Theoretically, it seems that the scaling form (\ref{shscaling}) is an unambiguous prediction of the theroy of critical phenomena \cite{lawrie1} and
ought to be observed sufficiently close to the zero-field critical point. Lowest-Landau-level scaling, on the other hand, is to be expected in large
fields, in the neighbourhood of the upper critical field.  There is in principle no region where both scaling forms could be simultaneously valid
\cite{lawrie0}.  There is, however, no reliable means of estimating the largest field in which critical-point scaling ought to be observable or the
smallest field consistent with lowest-Landau-level scaling.  Calculations are somewhat simplified by the lowest-Landau-level approximation, and
scaling functions have been obtained by both perturbative \cite{ruggieri,wilkin,hikami} and nonperturbative \cite{tesanovic} methods.  In particular,
Te$\check{\rm s}$anovi\'{c} and Andreev \cite{tesanovic} have obtained scaling functions which agree quite well with experimental data for the specific heat
and magnetization of YBCO, though the fit is rather better in the case of the magnetization than the specific heat \cite{pierson3}.

For the critical-point scaling function, no theoretical estimate has been obtained (although some general consequences of scaling have been discussed
in \cite{friesen}), and the calculation of this scaling function is the object of the work
reported here.  The calculation is based on the Ginzburg-Landau-Wilson model of an isotropic superconductor.  Although the superconductors of
interest are anisotropic, layered materials, this seems to be a reasonable approximation in the case of YBCO.  More generally, in fact, it is the divergence
of the coherence length near a critical point which gives rise to characteristic critical phenomena.  To the extent that the critical behaviour is that of a
3-dimensional system, therefore, one might expect the universal scaling function for an isotropic system to be that observed in the asymptotic critical region.
We assume that the magnetic coupling is weak enough for fluctuations in the vector potential $\bbox{A}(\bbox{r})$ to be neglected.  In fact, it is only
in this approximation that critical behaviour is to be expected \cite{halperin}. One barrier to this calculation is that, in the low-field regime, all the the Landau
levels must be included, and the eigenfunctions are extremely inconvenient to deal with.  Here, we exploit an integral representation of the order-parameter
propagator \cite{lawrie1} in which the sum over Landau levels is carried out once and for all.

Our initial attempts to estimate the scaling function ${\cal C}(x)$ made use of perturbation theory, which yields accurate results for the critical exponents.
Unfortunately, perturbation theory does not yield a well-controlled approximation to the function ${\cal C}(x)$, because it gives a spurious singularity in the
neighbourhood of $T_{\mathrm{c}2}(B)$ and we have been unable to cure this problem satisfactorily by {\it ad hoc} methods.  The alternative we adopt
here is to study a generalized Ginzburg-Landau-Wilson superconductor having an order parameter with $N$ complex components and to make use of
the expansion in powers of $1/N$.  The scaling function must be extracted by means of a renormalization-group analysis.  Curiously, we have not found
in the literature a formulation of the renormalization group that is well adapted to the use of the $1/N$ expansion as a means of approximation.  We have
therefore developed a suitable formulation, which is presented in detail for the zero-field case in \cite{lawrie2}. The extension of the $1/N$ formalism and
the renormalization-group analysis to the case of a nonzero magnetic field are summarized in sections II - IV below.  The calculation of the
scaling function requires a numerical means of estimating several cumbersome integrals, and the techniques we have devised for doing this are described
in section V.  As has long been known in connection with the estimation of, the convergence of the $1/N$ expansion is very poor.  The next-to-leading order
calculations reported here do not yield meaningful results for small values of $N$ (in particular for the physically relevant value $N=1$), but for larger values
we obtain scaling functions which are qualitatively similar to that observed experimentally. These results are presented and discussed in section VI.

\section{The $1/N$ expansion}
We consider the Ginzburg-Landau-Wilson  theory for an isotropic superconductor with $N$ complex order-parameter components $\phi_i(\bbox{r})$
in a fixed, uniform magnetic field of strength $B_0$. It is defined by the effective reduced Hamiltonian density
\begin{equation}
{\cal H} = \sum^N_{i=1} \left\{\left\vert\left(\bbox{\nabla}-i\bbox{A}(\bbox{r})\right) \phi_i(\bbox{ r}) \right\vert^2+t_0 \left\vert
\phi_i( \bbox{r}) \right\vert^2\right\} + {\lambda_0 \over 4N} \left( \sum^N_{i=1}
\left\vert \phi_i( \bbox{r} ) \right\vert^2 \right)^2\,,
\end{equation}
where $t_0$ is taken to be linear in temperature ($t_0\propto T-T_0$), and the coupling strength $\lambda_0$ to be temperature-independent.
A convenient choice of gauge for the magnetic vector potential is $\bbox{A}(\bbox{r})=B_0(0,x,0)$, corresponding to a uniform field in the $z$
direction, and we have absorbed the charge of a Cooper pair into the magnitudes of $\bbox{A}$ and $B_0$. As explained in \cite{lawrie2}, a
standard integral transformation of the Hubbard-Stratonovich type allows us to express the partition function as a functional integral over an
auxiliary field $\Psi$,
\begin{eqnarray}
Z&=&{\cal N}\int\prod_{i=1}^N{\cal D}\phi_i{\cal D}\phi_i^*\exp\left[-\int d^3r\,{\cal H}\right]\\
&=&\int{\cal D}\Psi\,\exp\left[-NH_{\mathrm{eff}}(\Psi)\right]\,,
\label{pf}
\end{eqnarray}
where the effective Hamiltonian is
\begin{equation}
H_{\mathrm{eff}}(\Psi)=\int d^3r\,\lambda_0^{-1}\Psi^2(\bbox{r})-{\mathrm Tr}_{\bbox{r},\bbox{r}'}\ln\Delta(\bbox{r},\bbox{r}';\Psi)
\end{equation}
and ${\cal N}$ is an irrelevant normalization constant. The propagator $\Delta(\bbox{r},\bbox{r}',\Psi)$ is the solution of
\begin{equation}
\left[-\nabla^2+2iB_0x\partial_y+B_0^2x^2+t_0+i\Psi(\bbox{r})\right]\Delta(\bbox{r},\bbox{r}';\Psi)=\delta(\bbox{r}-\bbox{r}')\,.
\label{ddeltaeqdelta}
\end{equation}
As in \cite{lawrie2}, it is helpful to formulate the $1/N$ expansion in terms of a self-energy $\tilde{t}_0$, which can be defined as follows.
The full 2-point function
\begin{equation}
G^{(2)}(\bbox{r},\bbox{r}')=\langle\phi_1^*(\bbox{r})\phi_1(\bbox{r}')\rangle=Z^{-1}\int{\cal D}\Psi\,\Delta(\bbox{r},\bbox{r'};\Psi)
e^{-NH_{\mathrm{eff}}(\Psi)}
\label{g2def}
\end{equation}
can be expressed as
\begin{equation}
G^{(2)}(\bbox{r},\bbox{r}')=\int\frac{dk_zd\sigma}{(2\pi)^2}B_0\sum_n\frac{\chi_{k_z,\sigma,n}^{\phantom{*}}(\bbox{r})\chi_{k_z,\sigma,n}^*(\bbox{r}')}
{\Gamma^{(2)}(n, k_z)}\,,
\end{equation}
where the $\chi_{k_z,\sigma,n}^{\phantom{*}}(\bbox{r})$ are eigenfunctions of the differential operator in (\ref{ddeltaeqdelta}), whose eigenvalues
are the Landau levels $E(n, k_z)=k_z^2+(2n+1)B_0+t_0$, and we define
\begin{equation}
\tilde{t}_0=\Gamma^{(2)}(0,0)\,.
\label{ttildedef}
\end{equation}
In the limit $B_0\to 0$, this agrees with the definition adopted in \cite{lawrie2}.  Because the Landau eigenfunctions are extremely inconvenient to deal with,
we shall exploit the fact that $e^{-i(x+x')(y-y')B_0/2}G^{(2)}(\bbox{r},\bbox{r}')$ is a translationally invariant function to write
\begin{equation}
G^{(2)}(\bbox{r},\bbox{r}')=e^{i(x+x')(y-y')B_0/2} \int\frac{d^3k}{(2\pi)^3}\,e^{i\bbox{k}\cdot(\bbox{r}-\bbox{r}')}G^{(2)}(\bbox{k})\,.
\end{equation}
Using the eigenfunctions given in \cite{lawrie1}, we find
\begin{equation}
\left[\Gamma^{(2)}(0,k_z)\right]^{-1}=(\pi B_0)^{-1}\int dk_xdk_y\,e^{-(k_x^2+k_y^2)/B_0}G^{(2)}(\bbox{k})\,.
\end{equation}

Owing to the factors of $N$ multiplying $H_{\mathrm{eff}}(\Psi)$ in (\ref{pf}) and (\ref{g2def}), the $1/N$ expansion is generated by the method
of steepest descent.  We expand $\Psi$ about the the position-independent saddle point by writing
\begin{equation}
\Psi(\bbox{r})=i(t_0+B_0-\tilde{t}_0+N^{-1}\delta)+(2N)^{-1/2}\psi(\bbox{r})\,,
\label{saddlepoint}
\end{equation}
where $\delta$ is defined by the condition $\langle\psi(\bbox{r})\rangle=0$.  The propagator $\Delta(\bbox{r},\bbox{r}';\Psi)$ can be expanded as
$\Delta(\bbox{r},\bbox{r}';\Psi)=\Delta(\bbox{r},\bbox{r}')+\textrm{O}(N^{-1/2})$, where the leading term is the solution of
\begin{equation}
\left[-\nabla^2+2iB_0x\partial_y+B^2+\tilde{t}_0-B_0\right]\Delta(\bbox{r},\bbox{r}')=\delta(\bbox{r}-\bbox{r}')\,.
\end{equation}
In real space, the diagrammatic expansion is identical to that explained in \cite{lawrie2}, to which we refer the reader for details, except that the
propagators are modified by the presence of the magnetic field.  The $\phi$ propagator $\Delta(\bbox{r},\bbox{r}')$ is given by
\begin{equation}
\Delta(\bbox{r},\bbox{r}')=e^{i(x+x')(y-y')B_0/2} \int\frac{d^3k}{(2\pi)^3}\,e^{i\bbox{k}\cdot(\bbox{r}-\bbox{r}')}\Delta(\bbox{k})
\label{phiprop}
\end{equation}
where, as obtained in \cite{lawrie1}, $\Delta(\bbox{k})$ has the integral representation
\begin{equation}
\Delta(\bbox{k})=\int_0^\infty du\,(\cosh B_0u)^{-1}\exp\left[-(k_z^2+\tilde{t}_0-B_0)u-(k_x^2+k_y^2)\tau(u)\right]
\label{deltaofk}
\end{equation}
with $\tau(u)=B_0^{-1}\tanh B_0u$. The $\psi$ propagator $D(\bbox{r}-\bbox{r}')$ is translationally invariant. Its inverse is
\begin{equation}
D^{-1}(\bbox{r}-\bbox{r}')=\lambda_0^{-1}\delta(\bbox{r}-\bbox{r}')+\frac{1}{2}\Delta(\bbox{r},\bbox{r}')\Delta(\bbox{r}',\bbox{r})
\end{equation}
and its Fourier transform is
\begin{equation}
D(\bbox{k})=\left[\lambda_0^{-1}+ \Pi(\bbox{k})\right]^{-1}
\label{psiprop}
\end{equation}
where
\begin{eqnarray}
\Pi(\bbox{k})&=&\frac{1}{2}\int\frac{d^3k'}{(2\pi)^3}\,\Delta(\bbox{k}')\Delta(\bbox{k}'+\bbox{k})\nonumber\\
&=&\left(\frac{1}{4\pi}\right)^{3/2}\frac{B_0}{2}\int_0^\infty du\,du'\frac{(u+u')^{-1/2}}{\sinh B_0(u+u')}\nonumber\\
&&\qquad\qquad\quad\times\exp\left[-\frac{uu'}{u+u'}\,k_z^2-\frac{\tau(u)\tau(u')}{\tau(u)+\tau(u')}(k_x^2+k_y^2)-(u+u')(\tilde{t}_0-B_0)\right]\,.
\end{eqnarray}

To make use of this expansion, we need to determine the counterterm $\delta$ introduced in (\ref{saddlepoint}) and the relation
between the self-energy $\tilde{t}_0$ and the variables $t_0$, $\lambda_0$ and $B_0$ with which we started. Consider first the expansion
for the 2-point function $G^{(2)}(\bbox{r},\bbox{r}')$ shown in figure 1(a).  The first term is just $\Delta(\bbox{r},\bbox{r}')$, which contains the exact
self-energy $\tilde{t}_0$, and satisfies (\ref{ttildedef}) by itself.  Thus, the counterterm $\delta$ is required to cancel the one-loop
contribution at $n=k_z=0$ and we find
\begin{eqnarray}
\delta&=&\left.\frac{1}{2\pi B_0}\int dk_kdk_y\,e^{-(k_x^2+k_y^2)/B_0}\int\frac{d^3k'}{(2\pi)^3}\,
\Delta(\bbox{k}+\bbox{k}')D(\bbox{k}')\right\vert_{k_z=0}\nonumber\\
&=&\frac{1}{2}\int\frac{d^3k}{(2\pi)^3}\,\widehat{\Delta}(\bbox{k})D(\bbox{k})\,,
\label{deltares}
\end{eqnarray}
where
\begin{equation}
\widehat{\Delta}(\bbox{k})=\int_0^\infty du\,\exp\left[-(k_z^2+\tilde{t}_0)u-(2B_0)^{-1}\left(1-e^{-2B_0u}\right)(k_x^2+k_y^2)\right]\,.
\end{equation}
The requirement that $\langle\psi(\bbox{r})\rangle=0$ yields a constraint equation, which implicitly determines $\tilde{t}_0$. Figure 1(b)
shows the expansion of $\langle\psi(\bbox{r})\rangle$ to order $1/N$; the function $f$ is the coefficient of $\psi(\bbox{r})$ in $H_{\mathrm{eff}}$,
as given in \cite{lawrie2}. From this we obtain
\begin{equation}
t_0=\Phi_0(\tilde{t}_0, \lambda_0,B_0)\equiv\tilde{t}_0-B_0-\frac{\lambda_0}{2}\,\Delta+N^{-1}\left[\frac{\lambda_0}{4}A
-\delta[1+\lambda_0\Pi({\bf 0})]\right]\,
\label{bareconstraint}
\end{equation}
where
\begin{eqnarray}
\Delta(\tilde{t}_0,\lambda_0,B_0)&=&\int\frac{d^3k}{(2\pi)^3}\,\Delta(\bbox{k})=\left(\frac{1}{4\pi}\right)^{3/2}B_0\int_0^\infty du
\frac{\exp[-(\tilde{t}_0-B_0)u]}{u^{1/2}\sinh B_0u}\\
A(\tilde{t}_0,\lambda_0,B_0)&=&\int\frac{d^3k}{(2\pi)^3}\,\Delta_3(\bbox{k})D(\bbox{k})\,.
\label{adef}
\end{eqnarray}
The function $\Delta_3(\bbox{k})$ corresponds to the loop of three $\phi$ propagators in figure 1(b), and is defined by
\begin{equation}
\Delta_3(\bbox{k})=\int d^3r'\,d^3r''\,e^{i\bbox{k}\cdot(\bbox{r}'-\bbox{r}'')}\Delta(\bbox{r},\bbox{r}')
\Delta(\bbox{r}',\bbox{r}'')\Delta(\bbox{r}'',\bbox{r})\,.
\label{delta3def}
\end{equation}
Straightforward but tedious algebra suffices to show that it is independent of $\bbox{r}$ and is given by
\begin{equation}
\Delta_3(\bbox{k})=-\frac{\partial\Pi(\bbox{k})}{\partial \tilde{t}_0}\,.
\end{equation}

Our aim is to investigate the scaling properties of the specific heat.  Within the Ginzburg-Landau-Wilson approximation, the specific
heat per unit volume per order-parameter component is given by
\begin{equation}
C=\frac{1}{2NV}\frac{\partial^2\ln Z}{\partial t_0^2}=(2N)^{-1}\int d^3r\sum_{i,j}\,
\langle\vert\phi_i(\bbox{0})\vert^2\,\vert\phi_j(\bbox{r})\vert^2\rangle_{\mathrm{c}}\,,
\end{equation}
where $V=\int d^3r$ is the volume and $\langle\ldots\rangle_{\mathrm{c}}$ denotes the connected correlation function. This correlation
function can be obtained directly as
\begin{equation}
\langle\vert\phi_i(\bbox{0})\vert^2\,\vert\phi_j(\bbox{r})\vert^2\rangle_{\mathrm{c}}=\delta_{ij}Z^{-1}\int{\cal D}\Psi\,
\Delta(\bbox{0},\bbox{r};\Psi)\Delta(\bbox{r},\bbox{0};\Psi)e^{-NH_{\mathrm{eff}}}\,,
\end{equation}
but it is not hard to obtain the convenient expression
\begin{equation}
C=\lambda_0^{-1}\left[1-\lambda_0^{-1}{\cal D}(\bbox{0})\right]
\label{barespht}
\end{equation}
where ${\cal D}(\bbox{k})=D(\bbox{k})+\textrm{O}(N^{-1})$ is the Fourier transform of the 2-point function $\langle\psi(\bbox{r})\psi(\bbox{r}')\rangle$.
To order $1/N$, this 2-point function is given by the sum of diagrams shown in figure 2, and is conveniently expressed in terms of a self-energy
$\Pi_\psi(\bbox{k})$ as
\begin{equation}
{\cal D}(\bbox{k})=D(\bbox{k})+N^{-1}D(\bbox{k})\Pi_\psi(\bbox{k})D(\bbox{k})\,.
\end{equation}
As explained in appendix A, the self-energy at $\bbox{k}=0$ is given by
\begin{equation}
\Pi_\psi(\bbox{0})=-\frac{1}{4}\frac{\partial A}{\partial\tilde{t}_0}
+\delta\frac{\partial \Pi(\bbox{0})}{\partial\tilde{t}_0}=-\frac{\partial}{\partial\tilde{t}_0}\left[\frac{1}{4}A-\delta\Pi(\bbox{0})\right]-\frac{\partial\delta}
{\partial\tilde{t}_0}\Pi(\bbox{0})\,,
\label{barepipsi}
\end{equation}
the second expression being convenient for the purpose of renormalization.

\section{Renormalization}

The scaling behaviour of thermodynamic functions emerges in the usual way from a renormalization-group analysis, but in
the context of the $1/N$ expansion this requires a nonstandard renormalization scheme, which is developed in detail
in \cite{lawrie2} for the theory with $B_0=0$.  According to this scheme, renormalized variables $\tilde{t}$, $t$, $z$ and $B$ are defined by
\begin{eqnarray}
\tilde{t}_0&=&\mu^2\tilde{t}\left[1-N^{-1}\frac{S_3}{6b}\ln z+\mathrm{O}(N^{-2})\right]\label{ttilderenorm}\\
t_0&=&t_{0\mathrm{c}}+\mu^2t\frac{(z+2a)}{z}\left[1-N^{-1}\frac{2S_3}{3b}\ln z+\mathrm{O}(N^{-2})\right]\label{trenorm}\\
\lambda_0^{-1}&=&\mu^{-1}z\left[1+N^{-1}\frac{4S_3}{3b}\ln z+\mathrm{O}(N^{-2})\right]\label{zrenorm}\\
B_0&=&\mu^2B\,.\label{brenorm}
\end{eqnarray}
In these expressions, $t_{0\mathrm{c}}$ is the value of $t_0$ at the zero-field critical point, $S_3=(2\pi)^{-2}$ is the usual factor arising from angular
integrations, and the constants $a=1/16\pi$ and $b=1/16$ arise from the large- and small-momentum limits of $\Pi(\bbox{k})$ when $B=0$.
As usual, $\mu$ is an arbitrary renormalization scale, with the dimensions of inverse length. The magnetic field requires no
renormalization;  the definition (\ref{brenorm}) simply serves to make $B$ dimensionless, as are $\tilde{t}$, $t$ and $z$. In this scheme,
critical behaviour is governed by an infrared-stable renormalization-group fixed point at $z=0$.  The criterion for renormalization is that renormalized
thermodynamic functions should have finite, non-zero limits as $z\to 0$, and we have implemented this requirement by a `minimal subtraction' of the
leading singularities proportional to $\ln z$.  It is crucial to our analysis that, as in the perturbative renormalization of \cite{lawrie1}, the presence of
a magnetic field introduces no additional divergences beyond those encountered at $B=0$ and we shall return to this point shortly.

For our immediate purposes, we need renormalized versions of the constraint equation (\ref{bareconstraint}) and the specific heat (\ref{barespht}).
The various integrals and subintegrals from which these are constructed must be re-expressed in terms of the renormalized variables. To this end, it
is convenient to introduce the dimensionless quantities
\begin{eqnarray}
\Pi_{\mathrm{R}}(\bbox{p};\alpha)&=&(\mu^2 B)^{1/2}\Pi(\bbox{k};\mu^2\tilde{t},\mu^2 B)\\
D_{\mathrm{R}}(\bbox{p};\alpha,z,B)&=&[z+B^{-1/2}\Pi_{\mathrm{R}}(\bbox{p};\alpha)]^{-1}\\
\Delta_{\mathrm{R}}(\alpha)&=&(\mu^2B)^{-1/2}\left[\Delta(\mu^2\tilde{t},\mu^2B)-\Delta(0,0)\right]\\
\widehat{\Delta}_{\mathrm{R}}(\bbox{p};\alpha)&=&\mu^2B\widehat{\Delta}(\bbox{k};\mu^2\tilde{t},\mu^2B)\\
A_{\mathrm{R}}(\alpha,z,B)&=&\mu^{-1}\left[A(\mu^2\tilde{t},\mu z^{-1}, \mu^2B)-4\delta(\mu^2\tilde{t},\mu z^{-1},\mu^2B)\Pi(\bbox{0};\mu^2\tilde{t},
\mu^2B)\right]\label{ardef}\\
\delta_{\mathrm{R}}(\alpha,z,B)&=&(\mu^4B)^{-1/2}\left[\delta(\mu^2\tilde{t},\mu z^{-1},\mu^2B)-\delta(0,\mu z^{-1},0)\right]\label{deltardef}\\
\Delta_{3\mathrm{R}}(\bbox{p};\alpha)&=&(\mu^2B)^{3/2}\Delta_3(\bbox{k};\mu^2\tilde{t},\mu^2B)\,,
\end{eqnarray}
with rescaled variables $\bbox{p}$ and $\alpha$ defined by
\begin{equation}
\bbox{p}=(\mu^2B)^{-1/2}\bbox{k}\qquad\textrm{and}\qquad\alpha=\tilde{t}/B\,.
\end{equation}
Subsequently, it will also be helpful to write
\begin{equation}
p_z^2=p^2\cos^2\theta\,,\qquad p_x^2+p_y^2=p^2\sin^2\theta\,.
\label{pandtheta}
\end{equation}

With this notation, the constraint equation becomes
\begin{eqnarray}
t&=&(z+2a)^{-1}\Phi(\tilde{t}, z, B)\\
\Phi(\tilde{t},z,B)&=&z(\tilde{t}-B)-\frac{1}{2}B^{1/2}\Delta_{\mathrm{R}}\nonumber\\
&&\quad+N^{-1}\left[\frac{1}{4}A_{\mathrm{R}}
+B^{1/2}\left(\Delta_{\mathrm{R}}+\frac{\alpha}{4}\frac{\partial\Delta_{\mathrm{R}}}{\partial\alpha}\right)\frac{S_3}{3b}\ln z\right.\nonumber\\
&&\qquad\qquad\left.-zB^{1/2}\delta_{\mathrm{R}}+\left(\frac{1}{2}\tilde{t}-\frac{2}{3}B\right)\frac{S_3}{b}z\ln z\right]+\mathrm{O}(N^{-2})\,.
\label{renormconstraint}
\end{eqnarray}
The dimensionless, renormalized specific heat $C_{\mathrm{R}}(\tilde{t}, z, B)$ is defined by
\begin{equation}
C(\tilde{t}_0,\lambda_0, B_0)=C(0,\lambda_0,0)+C_1\lambda_0^{-3}(t_0-t_{0\mathrm{c}})+\bar{Z}_t^{-2}C_{\mathrm{R}}(\tilde{t},z,B)\,,
\label{sphtrenorm}
\end{equation}
where
\begin{equation}
\bar{Z}_t=\frac{(z+2a)}{z}\left[1-N^{-1}\frac{2S_3}{3b}\ln z+\mathrm{O}(N^{-2})\right]
\end{equation}
is the renormalization factor appearing in (\ref{trenorm}). The dimensionless constant $C_1$ multiplies a non-singular term, whose presence
in three dimensions was first noted by Abe and Hikamai \cite{abe} and whose role in our renormalization scheme is discussed in \cite{lawrie2}.
Writing $C_{\mathrm{R}}(\tilde{t},z,B)=(z+2a)^2\bar{C}_{\mathrm{R}}(\tilde{t},z,B)$, we find
\begin{eqnarray}
\bar{C}_{\mathrm{R}}(\tilde{t},z,B)&=&-D_{\mathrm{R}}(\bbox{0};\alpha,z,B)-\frac{1}{2}C_1\Phi(\tilde{t},z,B)\nonumber\\
&&\quad+N^{-1}\left[E_1(\tilde{t},z,B)+E_2(\tilde{t},z,B)+E_3(\tilde{t},z,B)\right]+\mathrm{O}(N^{-2})\,,
\label{renormspht}
\end{eqnarray}
where
\begin{eqnarray}
E_1&=&\frac{1}{4}D_{\mathrm{R}}^2B^{-1}\frac{\partial A_{\mathrm{R}}}{\partial\alpha}-D_{\mathrm{R}}\frac{5S_3}{6b}\ln z -D_{\mathrm{R}}^2B^{-1/2}
\alpha\frac{\partial\Pi_{\mathrm{R}}}{\partial\alpha}\frac{S_3}{6b}\ln z\\
E_2&=&D_{\mathrm{R}}^2B^{-1}\Pi_{\mathrm{R}}\frac{\partial\delta_{\mathrm{R}}}{\partial\alpha}-\frac{S_3}{b}z^{-1}-D_{\mathrm{R}}\frac{S_3}{2b}\ln z
-C_1\left(z(\tilde{t}-B)-\frac{1}{2}B^{1/2}\Delta_{\mathrm{R}}\right)\frac{S_3}{b}\ln z\\
E_3&=&D_{\mathrm{R}}^2\frac{4S_3}{3b}z\ln z
\end{eqnarray}
and $D_{\mathrm{R}}$ and $\Pi_{\mathrm{R}}$ stand for $D_{\mathrm{R}}(\bbox{0};\alpha,z,B)$ and $\Pi_{\mathrm{R}}(\bbox{0};\alpha)$ respectively.
The integrals $A_{\mathrm{R}}(\alpha,z,B)$ and $\delta_{\mathrm{R}}(\alpha,z,B)$ defined in (\ref{ardef}) and (\ref{deltardef}) are both singular when
$z\to 0$, but each of the quantities $\Phi(\tilde{t},z,B)$ and $E_i(\tilde{t},z,B)$ has a finite limit, provided that we choose $C_1=-2/b^2$.  To verify this
assertion is not an entirely trivial matter.  In particular, to verify that the expression (\ref{renormconstraint}) for $\Phi(\tilde{t}, z, B)$ has a finite limit,
it is necessary to show that
\begin{equation}
A_{\mathrm{R}}(\alpha,z,B)=-B^{1/2}\left(4\Delta_{\mathrm{R}}(\alpha)+\alpha\frac{\partial\Delta_{\mathrm{R}}(\alpha)}{\partial\alpha}\right)
\frac{S_3}{3b}\ln z+\mathrm{O}(z\ln z)\,.
\end{equation}
Because the singularities arise from the large-$p$ region of integration, the required cancellations can be verified by means of large-$p$ expansions
of the subintegrals $\Pi_{\mathrm{R}}(\bbox{p};\alpha)$, $\Delta_{3\mathrm{R}}(\bbox{p};\alpha)$ and $\widehat{\Delta}_{\mathrm{R}}(\bbox{p};\alpha)$,
which are discussed in appendix B.

\section{Renormalization group and scaling}

The fact that the unrenormalized theory is independent of the renormalization scale $\mu$ leads in the standard way to renormalization-group
equations for the renormalized quantities $t(\tilde{t}, z, B)$ and $C_{\mathrm{R}}(\tilde{t}, z, B)$, which take the form
\begin{eqnarray}
\left[\beta(z)\frac{\partial}{\partial z}-(2-\eta(z))\tilde{t}\frac{\partial}{\partial\tilde{t}}-2B\frac{\partial}{\partial B}+\frac{1}{\nu(z)}\right]t(\tilde{t},z,B)&=&0\\
\left[\beta(z)\frac{\partial}{\partial z}-(2-\eta(z))\tilde{t}\frac{\partial}{\partial\tilde{t}}-2B\frac{\partial}{\partial B}
-\frac{\alpha(z)}{\nu(z)}\right]C_{\mathrm{R}}(\tilde{t},z,B)&=&0\,,\label{crg}
\end{eqnarray}
where $\alpha(z)=2-3\nu(z)$ and the remaining functions are those derived in \cite{lawrie2}. In contrast to perturbative renormalization schemes, the
additive renormalizations of the specific heat in (\ref{sphtrenorm}) are independent of $\mu$, so the associated renormalization group equation (\ref{crg})
is homogeneous.  The asymptotic critical behaviour with which we are concerned here is governed by the infrared-stable fixed point at $z=0$, where
$\beta(0)=0$ and the other function reduce to the critical exponents
\begin{equation}
\nu=1-\frac{16}{3\pi^2N}+\mathrm{O}(N^{-2}),\quad \alpha=-1+\frac{16}{\pi^2N}+\mathrm{O}(N^{-2}),\quad
\eta=\frac{4}{3\pi^2N}+\mathrm{O}(N^{-2}).
\label{exponents}
\end{equation}
For $z=0$, the renormalization-group equations are equivalent to the relations
\begin{eqnarray}
t(\tilde{t},0,B)&=&\ell^{1/\nu}t(\ell^{-(2-\eta)}\tilde{t},0,\ell^{-2}B)=\ell^{1/\nu}(2a)^{-1}\Phi(\ell^{-(2-\eta)}\tilde{t},0,\ell^{-2}B)\label{rgt}\\
C_{\mathrm{R}}(\tilde{t},0,B)&=&\ell^{-\alpha/\nu}C_{\mathrm{R}}(\ell^{-(2-\eta)}\tilde{t},0,\ell^{-2}B)\,,\label{rgc}
\end{eqnarray}
where $\ell$ is an arbitrary scaling factor.  The functions $t(\tilde{t},0,B)$ and $C_{\mathrm{R}}(\tilde{t},0,B)$ have infrared singularities when their first
argument $\tilde{t}$ vanishes. On the right-hand sides of (\ref{rgt}) and (\ref{rgc}), we exponentiate these singularities into the prefactors by choosing
$\ell$ to satisfy $\ell^{-(2-\eta)}\tilde{t}=1$.  Then, by setting $\ell=B^{1/2}L$, we find that $C_{\mathrm{R}}$ has the scaling form
\begin{equation}
C_{\mathrm{R}}=B^{-\alpha/2\nu}{\cal C}(tB^{-1/2\nu})
\end{equation}
where, with $x=tB^{-1/2\nu}$, the scaling function is
\begin{equation}
{\cal C}(x)=L^{-\alpha/\nu}(x)C_{\mathrm{R}}(1,0,L^{-2}(x))\,,
\label{finalc}
\end{equation}
the function $L(x)$ being determined by the constraint equation
\begin{equation}
2ax=L^{1/\nu}(x)\Phi(1,0,L^{-2}(x))\,.
\label{finalconstr}
\end{equation}
To obtain a numerical estimate of the scaling function ${\cal C}(x)$, we need an approximate means of evaluating the integrals in (\ref{renormconstraint})
and (\ref{renormspht}) which is discussed in the following section.

\section{Numerical estimation of integrals}

In order to determine the functions $C_{\mathrm R}(1, 0, L^{-2})$ and $\Phi(1, 0, L^{-2})$, and hence the scaling function ${\cal C}(x)$, we need to estimate
the renormalized counterparts of integrals such as (\ref{deltares}) and (\ref{adef}).  This requires analytic approximations to the functions
$\Pi_{\mathrm{R}}(\bbox{p}; \alpha)$, $\widehat{\Delta}_{\mathrm{R}}(\bbox{p}; \alpha)$ and $\Delta_{3\mathrm{R}}(\bbox{p}; \alpha)$, which are themselves
defined by rather intractable integrals.  This section indicates the methods of approximation we have used, focussing on the example of
$\Delta_{3\mathrm{R}}(\bbox{p}; \alpha)$, which we express in terms of the variables (\ref{pandtheta}) as $\Delta_{3\mathrm{R}}(p,\theta; \alpha)$. Having used
the renormalization group to replace $\tilde{t}$ with 1 and $B$ with $L^{-2}$ in (\ref{finalc}) and (\ref{finalconstr}), we have $\alpha=\tilde{t}/B=L^2$.

For large values of $p$, we approximate all of the subintegrals by means of the large-momentum expansions developed in appendix B.  Numerically, this
turns out to be a good approximation for $p\ge 6$.  In particular, this strategy allows us to cancel analytically the divergences that arise at the fixed point
$z=0$.

For $p<6$, an expansion in inverse powers of $\alpha$ is possible when $\alpha$ is large enough.  More specifically, in the case of $\Delta_{3\mathrm{R}}$,
we have an expansion of the form
\begin{equation}
\Delta_{3\mathrm{R}}(p, \theta; \alpha)=\alpha^{-3/2}\left[f_0(q,\theta)+\alpha^{-1}f_1(q,\theta)+\alpha^{-2}f_2(q,\theta)+\textrm{O}(\alpha^{-3})\right]
\end{equation}
where $q=\alpha^{-1/2}p$.  (This entails, of course, a rescaling of the integration variable in the final integral (\ref{adef}).)  Using the representation
(\ref{delta3vrep}), the change of variables $v=w/\alpha$, $v'=w'/\alpha$ leads to a power series expansion in which each of the remaining integrals
can be calculated analytically.  We find that the functions $f_i(q,\theta)$ are given by
\begin{eqnarray}
f_0&=&(8\pi)^{-1}Q\,,\qquad f_1=(16\pi)^{-1}(Q+8Q^2)\nonumber\\
f_2&=&(96\pi)^{-1}[3Q+16Q^2+(128+4q^2s^2)Q^3+96q^2s^2Q^4]\,,
\end{eqnarray}
where $Q=(q^2+4)^{-1}$ and $s=\sin\theta$.  In practice, we have used this approximation for $\alpha>2.25$, where it appears to yield results of
satisfactory accuracy.

For $p<6$ and $\alpha<2.25$ no systematic expansion in any small parameter will serve our purpose. Instead, we have devised an approximation
scheme which we again illustrate for the example of $\Delta_{3\mathrm{R}}$.  The basic strategy is to evaluate the double integral (\ref{delta3r}) numerically
for selected values of $p$, $\theta$ and $\alpha$ and to construct an interpolating function from these numerical values.  To interpolate simultaneously in all
three variables is a difficult undertaking, however.  To simplify it, we introduce a further approximation, which reduces the function of three variables
to several functions, each depending on only two variables.  In the expression (\ref{delta3r}), we make the change of integration variables
\begin{equation}
u=\rho\cos^2\phi\,,\qquad u'=\rho\sin^2\phi\,.
\end{equation}
The integral becomes (again, with the notation $s=\sin\theta$ and $c=\cos\theta$)
\begin{eqnarray}
\Delta_{3\mathrm{R}}(p,\theta;\alpha)&=&\frac{1}{2(4\pi)^{3/2}}\int_0^{\pi/2}d\phi\,\sin(2\phi)\int_0^\infty d\rho\,\frac{\rho^{3/2}}{\sinh\rho}\nonumber\\
&&\qquad\qquad\qquad\times \exp\left[-(\alpha-1)p-(p^2c^2/4)\rho\sin^2(2\phi)-p^2s^2{\cal T}(\rho,\phi)\right]\,,\\
{\cal T}(\rho,\phi)&=&\frac{\tanh(\rho\cos^2\phi)\tanh(\rho\sin^2\phi)}{\tanh(\rho\cos^2\phi)+\tanh(\rho\sin^2\phi)}\,.
\end{eqnarray}
Our approximation scheme is based on the observation that ${\cal T}(\rho,\phi)\approx(\rho/4)\sin^2(2\phi)$ for $\rho\to 0$ while
${\cal T}(\rho,\phi)\approx \frac{1}{2}$ for $\rho\to\infty$, except at the endpoints $\phi=0$ and $\phi=\pi/2$.  We divide the region of integration into two parts:
region I, where $0<\rho<S(\phi)$, and region II, where $S(\phi)<\rho<\infty$.  The boundary $\rho=S(\phi)$ is determined in a manner to be explained
shortly.  We have $\Delta_{3\mathrm{R}}=\Delta_{3\mathrm{R}}^{\mathrm{I}}+\Delta_{3\mathrm{R}}^{\mathrm{II}}$, where
\begin{eqnarray}
\Delta_{3\mathrm{R}}^{\mathrm{I}}&=&\frac{1}{2(4\pi)^{3/2}}\int_0^{\pi/2}d\phi\,\sin(2\phi)\int_0^{S(\phi)} d\rho\,\frac{\rho^{3/2}}{\sinh\rho}\nonumber\\
&&\qquad\qquad\qquad\times\exp\left[-(\alpha-1)\rho-(p^2\rho/4)\sin^2(2\phi)+p^2s^2X^{\mathrm{I}}\right] \label{deltai}\\
\Delta_{3\mathrm{R}}^{\mathrm{II}}&=&\frac{e^{-p^2/2}}{2(4\pi)^{3/2}}\int_0^{\pi/2}d\phi\,\sin(2\phi)\int_{S(\phi)}^\infty d\rho\,\frac{\rho^{3/2}}{\sinh\rho}\nonumber\\
&&\qquad\qquad\qquad\times\exp\left[-(\alpha-1)\rho-(p^2c^2/4)(\rho\sin^2(2\phi)-2)+p^2s^2X^{\mathrm{II}}\right] \label{deltaii}
\end{eqnarray}
with
\begin{eqnarray}
X^{\mathrm{I}}(\rho,\phi)&=&\left[(\rho/4)\sin^2(2\phi)-{\cal T}(\rho,\phi)\right]\\
X^{\mathrm{II}}(\rho,\phi)&=&\left[{\textstyle\frac{1}{2}}-{\cal T}(\rho,\phi)\right]
\end{eqnarray}
and we propose to expand the integrands of (\ref{deltai}) and (\ref{deltaii}) in powers of $X^{\mathrm{I}}$ and $X^{\mathrm{II}}$ respectively. The boundary
$\rho=S(\phi)$ is chosen as the locus on which $X^{\mathrm{I}}=X^{\mathrm{II}}$, namely $S(\phi)=2/\sin^2(2\phi)$, so that the two expansions match
term by term on the boundary. With this choice, $X^{\mathrm{I}}(\rho,\phi)$ and $X^{\mathrm{II}}(\rho,\phi)$ are always smaller than the boundary
value $X^S(\phi)=\frac{1}{2}-{\cal T}(S(\phi),\phi)$, which itself has a maximum value of approximately 0.18 at $\phi=0$ and $\phi=\pi/2$.  Moreover,
the quantities $p^2s^2X^A$ are positive, so the expansion of each integrand converges monotonically.  This is not necessarily true of the integrals, but in
practice we have found that retaining only the first two terms of each expansion yields results that are fairly accurate and match smoothly to the large-$\alpha$
and large-$p$ expansions.  It will be seen that each term in the expansion of $\Delta_{3\mathrm{R}}^{\mathrm{I}}(p,\theta;\alpha)$ is of the form
$(p^2s^2)^nf_n^{\mathrm{I}}(p^2,\alpha)$, while each term in the expansion of $\Delta_{3\mathrm{R}}^{\mathrm{II}}(p,\theta;\alpha)$ is of the form
$e^{-p^2/2}(p^2s^2)^nf_n^{\mathrm{II}}(p^2c^2,\alpha)$.  Each of the functions $f^A_n$ depends only on two variables.  We have obtained interpolations
for these functions, for $n=0,1$, giving final approximations for $\Delta^A_{3\mathrm{R}}$ of the form
\begin{eqnarray}
\Delta^{\mathrm{I}}_{3\mathrm{R}}(p,\theta;\alpha)&=&\left[\frac{R^{\mathrm{I}}_{1,0}(\alpha)}{1+\sum_{n=1}^6R_{1,n}(\alpha)p^{2n}}\right]^{1/6}
+p^2s^2\frac{R^{\mathrm{I}}_{2,0}(\alpha)}{\left[1+\sum_{n=1}^6R^{\mathrm{I}}_{2,n}(\alpha)p^{2n}\right]^{1/2}}\\
\Delta^{\mathrm{II}}_{3\mathrm{R}}(p,\theta;\alpha)&=&\frac{e^{-p^2/2}R^{\mathrm{II}}_{1,0}(\alpha)\left[1+R^{\mathrm{II}}_{1,1}(\alpha)p^2c^2\right]}
{1+ R^{\mathrm{II}}_{1,2}(\alpha)p^2c^2 +R^{\mathrm{II}}_{1,3}(\alpha)p^4c^4}
+p^2s^2\frac{e^{-p^2/2}R^{\mathrm{II}}_{2,0}(\alpha)}{ 1+R^{\mathrm{II}}_{2,1}(\alpha)p^2c^2}\,,
\end{eqnarray}
where the $R^A_{i,j}$ are rational approximants obtained from the Thiele interpolation formula. The form of these interpolating functions (and those for
$\Pi_{\mathrm{R}}(\bbox{p}; \alpha)$ and $\widehat{\Delta}_{\mathrm{R}}(\bbox{p}; \alpha)$, which we do not give explicitly)  is chosen so as to give the
correct behaviour at large values of $p$ and to allow the integrals over $\theta$ to be performed analytically in the final calculations of $A_{\mathrm{R}}$
and $\delta_{\mathrm{R}}$.

\section{Results and discussion}

As compared with perturbation theory (the expansion in powers of $\lambda_0$), the $1/N$ expansion has a key advantage when applied to the
problem of a superconductor in a magnetic field.  At the lowest order of perturbation theory, it is not hard to show that the specific heat behaves as
\begin{equation}
C\sim B_0(t_0-t_{0\mathrm{c}}+B_0)^{-3/2}\sim B_0\left[T-T_{\mathrm{c}2}(B_0)\right]^{-3/2}
\end{equation}
in the neighbourhood of the line $T=T_{\mathrm{c}2}(B_0)$. In terms of the formalism used in this paper, this approximation corresponds to taking
$C\approx\Pi(\bbox{0})$ in (\ref{barespht}) and $\tilde{t}_0\approx t_0-t_{0\mathrm{c}}+B_0$ in the constraint equation (\ref{bareconstraint}).  This
divergence is entirely spurious.  Experimentally, there is no sign of it and theoretically it can be removed by means of a self-energy resummation
of the Hartree variety.  The $1/N$ expansion incorporates this resummation in a way which allows a renormalization-group analysis of the scaling
behaviour to be systematically pursued.

On the other hand, the $1/N$ expansion has serious drawbacks.  Even at the next-to-leading order we have used here, calculations are extremely
cumbersome, and this is unfortunate, because the convergence of the expansion is notoriously poor.  With the relevant value $N=1$ for the number
of complex order-parameter components, the specific heat exponent given in (\ref{exponents}) is $\alpha\approx 0.62$; compared with the best
theoretical and experimental value for the XY model $\alpha_{\mathrm{XY}}\approx -0.013$, it is about fifty times too large and has the wrong sign!
For the correlation-length exponent, we have $\nu\approx0.46$ compared with $\nu_{\mathrm{XY}}\approx 0.67$.

Using the formalism and numerical methods summarized above, we have obtained estimates for the specific-heat scaling function ${\cal C}(x)$
as given in (\ref{finalc}).  Here too, we find that the convergence is poor;  for small values of $N$, the next-to-leading terms are larger than the
leading terms.  Matters are somewhat improved if the XY exponents $\nu_{\mathrm{XY}}$ and $\alpha_{\mathrm{XY}}$ are substituted in
(\ref{finalc}) and (\ref{finalconstr}) for those shown in (\ref{exponents}).  Here, we present only the best results (as judged by their qualitative
similarity to experimental data) that we have been able to obtain by this strategy.  Figure 3 shows the scaling function calculated for values of
$N$ between 10 and 20 and, for comparison, figure 4 reproduces the experimental data reported in \cite{overend}.  For more negative values of $x$
than those shown in figure 3, the calculated curves diverge rapidly, either to large positive values or to large negative values, and our approximations
are clearly inadequate in this region.  The reason for this is not entirely clear to us.  One possibility is that our neglect of the non-zero order parameter
$\langle\phi(\bbox{r})\rangle$ in the mixed state becomes seriously inadequate at temperatures a little below $T_{\mathrm{c}2}(B_0)$.  In the
lowest-Landau-level approximation, it seems to be possible to continue the scaling function of the homogeneous normal state to temperatures
well below $T_{\mathrm{c}2}(B_0)$, but the same may not be true of the critical-point scaling function.

Since we cannot obtain reliable results for the physically relevant number of order-parameter components $N=1$, a detailed fit of our calculated
scaling function to the data would have little meaning, and we have not attempted it.  For larger values of $N$, it is clear that the calculated
scaling function does reproduce the qualitative features of the experimentally determined function in the region where our approximations appear
to work. Tentatively, at least, it seems reasonable to conclude that the critical-point scaling implied by the Ginzburg-Landau-Wilson model is
consistent with what is actually observed.  A less tentative conclusion is that some much better method of approximation  than those currently
available is needed to test this scaling prediction quantitatively.

Whether the scaling observed in YBCO really corresponds to a regime dominated by critical-point fluctuations is another matter.  Indeed, the scaling
functions exhibited in figure 3 are qualitatively very similar to the 3-dimensional scaling function of the lowest-Landau-level approximation estimated by
Tes$\check{\rm a}$novi$\acute{\rm c}$ and Andreev \cite{tesanovic}. It is far from clear, therefore, that a quantitatively more reliable estimate of
the critical-point scaling function, should it be obtainable, would serve to discriminate between the two scaling hypotheses.

In this paper, we have attempted to estimate the scaling function associated with asymptotic critical behaviour, which is controlled by the
renormalization-group fixed point $z=0$.  In principle, the formalism and numerical approximations described here should also facilitate
an investigation of the competition between low-field critical-point scaling and high-field lowest-Landau-level scaling in the intermediate region where
neither type of scaling is exactly valid.  We plan to address this issue in a future publication.

\section*{Acknowledgments}

The financial support of the Engineering and Physical Science Research Council  through Research Grant GR/M09261 is gratefully acknowledged.
D.J.L. wishes to thank Alan Dorsey for helpful discussions.

\appendix
\section{Calculation of vertex functions}
As explained in detail in \cite{lawrie2}, the basic elements of Feynman diagrams in the $1/N$ expansion are vertex functions of the form
\begin{equation}
\Delta_n(\bbox{r}_1,\ldots,\bbox{r_n})=\Delta(\bbox{r}_1,\bbox{r}_2)\Delta(\bbox{r}_2,\bbox{r}_3)\ldots\Delta(\bbox{r}_n,\bbox{r}_1)\,.
\end{equation}
These are gauge-invariant functions, and therefore also translationally invariant, but in the presence of a magnetic field, the form of the
propagator (\ref{phiprop}) makes them somewhat awkward to handle.  For the purposes of this paper, the fact that we usually need to
integrate over one or more of the arguments $\bbox{r}_i$ leads to some simplification. Let us write
\begin{equation}
\Delta(\bbox{r},\bbox{r}')=e^{i(x+x')(y-y')B_0/2}\bar{\Delta}(\bbox{r}-\bbox{r}')
\end{equation}
where $\bar{\Delta}(\bbox{r})$ is the function whose Fourier transform is given in (\ref{deltaofk}). We find that
\begin{equation}
\int d^3r''\,\Delta(\bbox{r},\bbox{r}'')\Delta(\bbox{r}'',\bbox{r}')=e^{i(x+x')(y-y')B_0/2}\bar{\Delta}_2(\bbox{r}-\bbox{r}')\,,
\end{equation}
with
\begin{equation}
\bar{\Delta}_2(\bbox{r})=\int d^3r'\,e^{i(x'y-y'x)B_0/2}\bar{\Delta}(\bbox{r}-\bbox{r}')\bar{\Delta}(\bbox{r}')\,.
\end{equation}
A lengthy, but straightforward calculation shows that the Fourier transform of $\bar{\Delta}_2(\bbox{r})$ is
\begin{equation}
\bar{\Delta}_2(\bbox{k})=-\frac{\partial\Delta(\bbox{k})}{\partial \tilde{t}_0}\,.
\end{equation}
When $B_0=0$, this reduces to the familiar fact that $\partial[(k^2+\tilde{t}_0)^{-1}]/\partial\tilde{t}_0=-(k^2+\tilde{t}_0)^{-2}$. The
function $\Delta_3(\bbox{k})$ defined in (\ref{delta3def}) is equivalent to
\begin{equation}
\Delta_3(\bbox{k})=\int\frac{d^3k'}{(2\pi)^3}\,\Delta(\bbox{k}')\bar{\Delta}_2(\bbox{k}+\bbox{k}')
=-\frac{1}{2}\frac{\partial}{\partial\tilde{t}_0}\int\frac{d^3k'}{(2\pi)^3}\,\Delta(\bbox{k}')\Delta(\bbox{k}+\bbox{k}')
=-\frac{\partial\Pi(\bbox{k})}{\partial\tilde{t}_0}\,.
\label{delta3}
\end{equation}
In the same way, we can define $\bar{\Delta}_3(\bbox{r}-\bbox{r}')$ by
\begin{equation}
\int d^3r''d^3r'''\,\Delta(\bbox{r},\bbox{r}'')\Delta(\bbox{r}'',\bbox{r}''')\Delta(\bbox{r}''',\bbox{r}')
=e^{i(x+x')(y-y')B_0/2}\bar{\Delta}_3(\bbox{r}-\bbox{r}')\,,
\end{equation}
and find that its Fourier transform is
\begin{equation}
\bar{\Delta}_3(\bbox{k})=-\frac{1}{2}\frac{\partial\bar{\Delta}_2(\bbox{k})}{\partial \tilde{t}_0}
=\frac{1}{2}\frac{\partial^2\Delta(\bbox{k})}{\partial \tilde{t}_0^2}\,.
\end{equation}

Consider now the self-energy diagrams shown in figure 2, which are to be evaluated with the external wavevector equal to
zero. The first one is
\begin{eqnarray}
\Pi_\psi^{(1)}(\bbox{0})&=&\int\frac{d^3k}{(2\pi)^3}\frac{d^3k'}{(2\pi)^3}\,D(\bbox{k})\bar{\Delta}_2(\bbox{k}')
\bar{\Delta}_2(\bbox{k}+\bbox{k}')\nonumber\\
&=&-\int\frac{d^3k}{(2\pi)^3}\frac{d^3k'}{(2\pi)^3}\,D(\bbox{k})\frac{\partial\Delta(\bbox{k})}{\partial\tilde{t}_0}
\bar{\Delta}_2(\bbox{k}+\bbox{k}')
\end{eqnarray}
and the second is
\begin{eqnarray}
\Pi_\psi^{(2)}(\bbox{0})&=&\int\frac{d^3k}{(2\pi)^3}\frac{d^3k'}{(2\pi)^3}\,D(\bbox{k})\Delta(\bbox{k}')
\bar{\Delta}_3(\bbox{k}+\bbox{k}')\nonumber\\
&=&-\frac{1}{2}\int\frac{d^3k}{(2\pi)^3}\frac{d^3k'}{(2\pi)^3}\,D(\bbox{k})\Delta(\bbox{k}')
\frac{\partial \bar{\Delta}_2(\bbox{k}+\bbox{k}')}{\partial\tilde{t}_0}\,
\end{eqnarray}
so we can use (\ref{delta3}) to write
\begin{equation}
\Pi_\psi^{(1)}(\bbox{0})+2\Pi_\psi^{(2)}(\bbox{0})=-\int\frac{d^3k}{(2\pi)^3}\,D(\bbox{k})\frac{\partial\Delta_3(\bbox{k})}{\partial\tilde{t}_0}\,.
\end{equation}
Using the expression (\ref{psiprop}) for $D(\bbox{k})$ and the first expression in (\ref{delta3}) for $\Delta_3(\bbox{k})$, we find that
\begin{equation}
\frac{\partial D(\bbox{k})}{\partial\tilde{t}_0}=-D(\bbox{k})^2\Delta_3(\bbox{k})\,,
\end{equation}
so the third diagram of figure 2 is
\begin{eqnarray}
\Pi_\psi^{(3)}(\bbox{0})&=&\int\frac{d^3k}{(2\pi)^3}\,\Delta_3(\bbox{k})D(\bbox{k})^2\Delta_3(\bbox{k})\nonumber\\
&=&-\int\frac{d^3k}{(2\pi)^3}\,\frac{\partial D(\bbox{k})}{\partial\tilde{t}_0}\,\Delta_3(\bbox{k})
\end{eqnarray}
and we obtain
\begin{equation}
\Pi_\psi^{(1)}(\bbox{0})+2\Pi_\psi^{(2)}(\bbox{0})-\Pi_\psi^{(3)}(\bbox{0})
=-\frac{\partial}{\partial\tilde{t}_0}\int\frac{d^3k}{(2\pi)^3}\, D(\bbox{k})\Delta_3(\bbox{k})
=-\frac{\partial A(\tilde{t}_0,\lambda_0,B_0)}{\partial\tilde{t}_0}\,.
\end{equation}
The final diagram in figure 2 is
\begin{equation}
\Pi_\psi^{(4)}(\bbox{0})=\Delta_3(\bbox{0})=-\frac{\partial\Pi(\bbox{0})}{\partial\tilde{t}_0}\,.
\end{equation}

\section{Large-momentum expansions}

To verify that the constraint equation and the specific heat can be correctly renormalized, and also to assist the numerical
estimation of the renormalized quantities, we require expansions of the subintegrals $\Pi_{\mathrm{R}}(\bbox{p};\alpha)$,
$\Delta_{3\mathrm{R}}(\bbox{p};\alpha)$ and $\widehat{\Delta}_{\mathrm{R}}(\bbox{p};\alpha)$. We use the notation
indicated in (\ref{pandtheta}) and the abbreviations $s=\sin\theta$ and $c=\cos\theta$.  For $\widehat{\Delta}_{\mathrm{R}}(\bbox{p};\alpha)$,
the expansion
\begin{eqnarray}
\widehat{\Delta}_{\mathrm{R}}(p,\theta;\alpha)&=&\int_0^\infty du\,\exp\left[-(p^2c^2+\alpha)u-\frac{1}{2}(1-e^{-2u})p^2s^2\right]\nonumber\\
&=&p^{-2}-(\alpha-2s^2)p^{-4}+\left[\alpha^2-(6\alpha+4)s^2+12s^4\right]p^{-6}+\mathrm{O}(p^{-8})
\end{eqnarray}
follows trivially from the change of variable $u=v/p^2$. For $\Delta_{3\mathrm{R}}(\bbox{p};\alpha)$ we have the expression
\begin{equation}
\Delta_{3\mathrm{R}}=\frac{1}{2(4\pi)^{3/2}}\int_0^\infty du\,du'\,\frac{\displaystyle (u+u')^{1/2}\exp\left[-(\alpha-1)(u+u')-\frac{uu'}{(u+u')}p^2c^2-\frac{\tau\tau'}{(\tau+\tau')}
p^2s^2\right]}{\sinh(u+u')}
\label{delta3r}
\end{equation}
where $\tau=\tanh u$ and $\tau'=\tanh u'$. By virtue of the symmetry of the integrand under interchange of $u$ and $u'$, the region of integration
$0\le u'\le u$ yields exactly half of the integral. In this region, we can make the change of variable
\begin{equation}
u+u'=v+v'\,,\qquad 4uu'/(u+u')=v'
\end{equation}
to obtain
\begin{equation}
\Delta_{3\mathrm{R}}=\frac{1}{4(4\pi)^{3/2}}\int_0^\infty dv\,dv'\,\frac{(v+v')\exp\left[-(\alpha-1)(v+v')-\frac{1}{4}v'p^2c^2-\sigma(v,v')p^2s^2\right]}
{v^{1/2}\sinh(v+v')}\,,
\label{delta3vrep}
\end{equation}
where
\begin{equation}
\sigma(v,v')=\left[\cosh(v+v')-\cosh\left(\sqrt{v(v+v')}\right)\right]/2\sinh(v+v')\,.
\end{equation}
A further change of variable $v'=v''/p^2$ facilitates an expansion in powers of $p^{-2}$, with a result of the form
\begin{eqnarray}
\Delta_{3\mathrm{R}}(p,\theta;\alpha)=Q_0(\alpha)p^{-2}+&&\left[Q_1(\alpha)+Q_2(\alpha)s^2\right]p^{-4}\nonumber\\
&&+\left[Q_3(\alpha)+Q_4(\alpha)s^2+Q_5(\alpha)s^4\right]p^{-6}+\mathrm{O}(p^{-8})\,.
\end{eqnarray}
The coefficients $Q_i(\alpha)$ are
\begin{equation}
Q_i(\alpha)=\left(\frac{1}{4\pi}\right)^{3/2}\int_0^\infty dv\frac{v^{1/2}\exp[-(\alpha-1)v]}{\sinh v}\,{\cal Q}_i(v,\alpha)
\label{qiints}
\end{equation}
with
\begin{eqnarray*}
{\cal Q}_0(v,\alpha)&=&1\,,\quad{\cal Q}_1(v,\alpha)=4(v^{-1}-\coth v+1-\alpha)\,,\quad{\cal Q}_2(v,\alpha)=2(\coth v-v^{-1})\\
{\cal Q}_3(v,\alpha)&=&16\left[\alpha^2-2\alpha+2(\coth v-1+\alpha)(\coth v-v^{-1})\right]\\
{\cal Q}_4(v,\alpha)&=&4\left[5+3(\coth v-v^{-1})(v^{-1}-4\coth v+2-2\alpha)\right]\\
{\cal Q}_5(v,\alpha)&=&12(\coth v-v^{-1})^2\,.
\end{eqnarray*}
The function $\Pi_{\mathrm{R}}(\bbox{p};\alpha)$ satisfies $\partial\Pi_{\mathrm{R}}(\bbox{p};\alpha)/\partial\alpha=-\Delta_{3\mathrm{R}}(\bbox{p};\alpha)$,
but does not itself have an expansion in powers of $p^{-2}$.  At $B=0$, we have the exact result
\begin{equation}
\Pi(\bbox{k};\tilde{t},0)=(8\pi k)^{-1}\tan^{-1}(k/2\tilde{t}^{1/2})=bk^{-1}-4a\tilde{t}^{1/2}k^{-2}+\mathrm{O}(k^{-4})\,,
\end{equation}
with $k=\vert\bbox{k}\vert$, which implies that $\Pi_{\mathrm{R}}(\bbox{p};\alpha)$ has the limiting form
\begin{equation}
\Pi_{\mathrm{R}}(\bbox{p};\alpha)=bp^{-1}-4a\alpha^{1/2}p^{-2}+\mathrm{O}(p^{-4})
\end{equation}
as $\alpha\to\infty$.  By integrating $\Delta_{3\mathrm{R}}(\bbox{p};\alpha)$ with this boundary condition, we obtain the expansion
\begin{equation}
\Pi_{\mathrm{R}}(\bbox{p};\alpha)=bp^{-1}+\Delta_{\mathrm{R}}(\alpha)p^{-2}+\mathrm{O}(p^{-4})\,,
\end{equation}
which is sufficient for our purposes.

In  the constraint equation (\ref{renormconstraint}) and the specific heat (\ref{renormspht}), singularities at $z\to 0$ arise from the large-$p$
region of integration in integrals of the form
\begin{equation}
\int\frac{d^3p}{(2\pi)^3}\,D_{\mathrm{R}}(\bbox{p};\alpha,z,B)f(\bbox{p};\alpha)\,.
\end{equation}
By restricting the range of integration to $\vert\bbox{p}\vert\ge p_0$, where the value of $p_0$ is immaterial, the leading singularities can be
extracted by means of the power-series expansions given above.  Using the expansion $D_{\mathrm{R}}^{-1}(\bbox{p};\alpha,z,B)
=z+B^{-1/2}bp^{-1}+B^{-1/2}\Delta_{\mathrm{R}}(\alpha)p^{-2}+\mathrm{O}(p^{-4})$, we encounter the three divergent integrals
\begin{eqnarray}
\int\frac{d^3p}{(2\pi)^3}\frac{1}{(z+B^{-1/2}bp^{-1})p^4}&=&-B^{1/2}\frac{S_{3}}{b}\,\ln z+\ldots\\
\int\frac{d^3p}{(2\pi)^3}\frac{1}{(z+B^{-1/2}bp^{-1})^2p^4}&=&B^{1/2}\frac{S_3}{b}\,z^{-1} +\ldots\\
\int\frac{d^3p}{(2\pi)^3}\frac{1}{(z+B^{-1/2}bp^{-1})^3p^6}&=&-B^{3/2}\frac{S_3}{b^3}\,\ln z +\ldots\,,
\end{eqnarray}
where the ellipsis represent less singular terms.  These results, together with straightforward, though tedious, manipulations of the integrals
(\ref{qiints}) suffice to verify that the functions $\Phi$, $E_1$ and $E_2$ have finite limits and, in (\ref{renormconstraint}), that
$z\delta_{\mathrm{R}}=-\Delta_{\mathrm{R}}S_3/2b+\ldots$.

%%%%%%%%%%%%%%%%%%%%%%%%%%%%%%%%%%%%%%%%
%%%%%%%%%%%%%%%%%%%%%%%%%%%%%%%%%%%%%%%%

\begin{figure}
\caption{Diagrammatic representation of ({\it a}) the order-parameter 2-point function and ({\it b}) the expectation value $\langle\psi(\bbox{r})\rangle$
at next-to leading order.}
\label{fig1}
\end{figure}
\begin{figure}
\caption{Diagrammatic representation of the 2-point function for the auxiliary field $\psi$ at next-to-leading order.}
\label{fig2}
\end{figure}
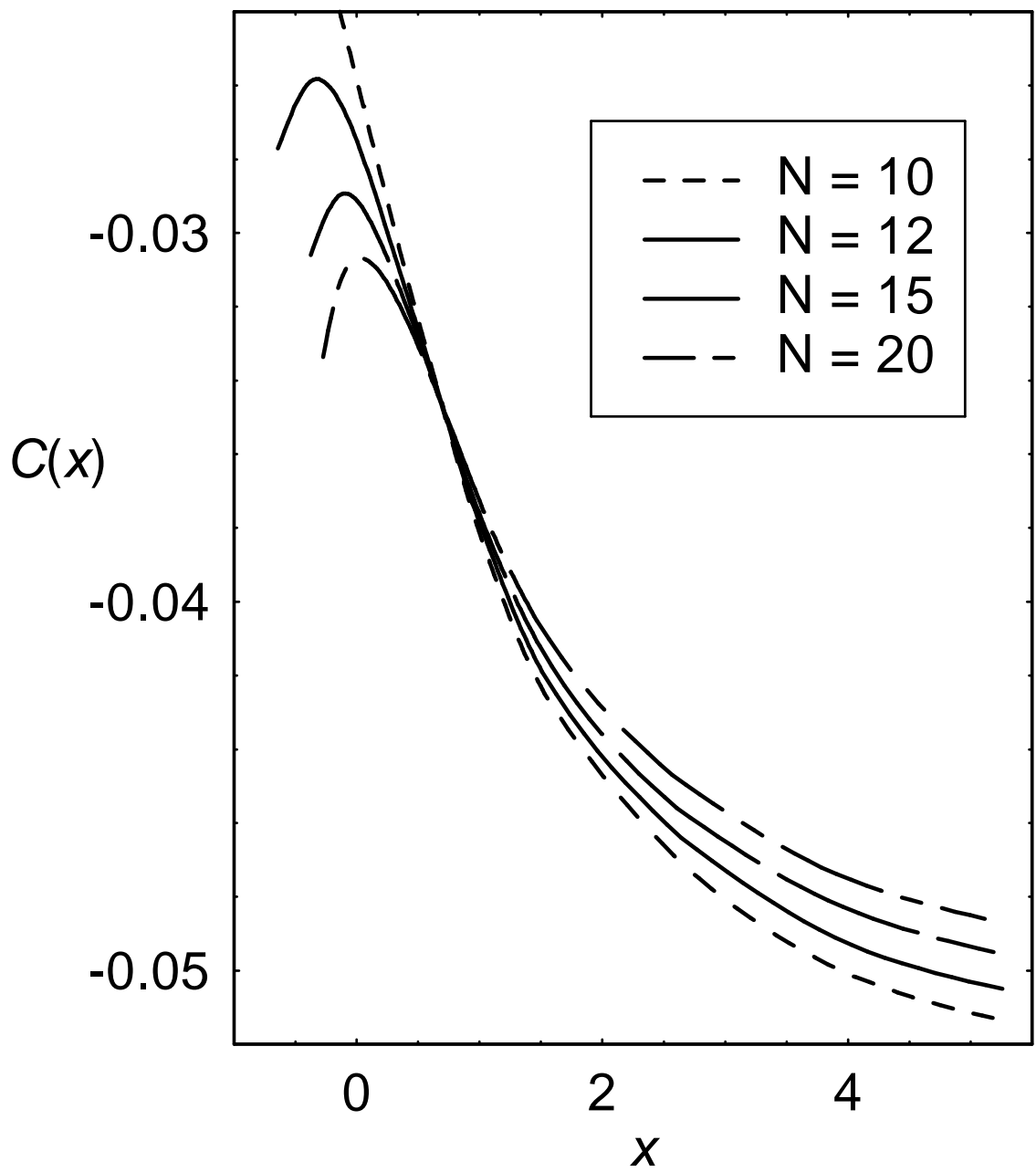
\begin{figure}
\caption{Numerical results for the specific-heat scaling function ${\cal C}(x)$ for several values of $N$.}
\label{fig3}
\end{figure}
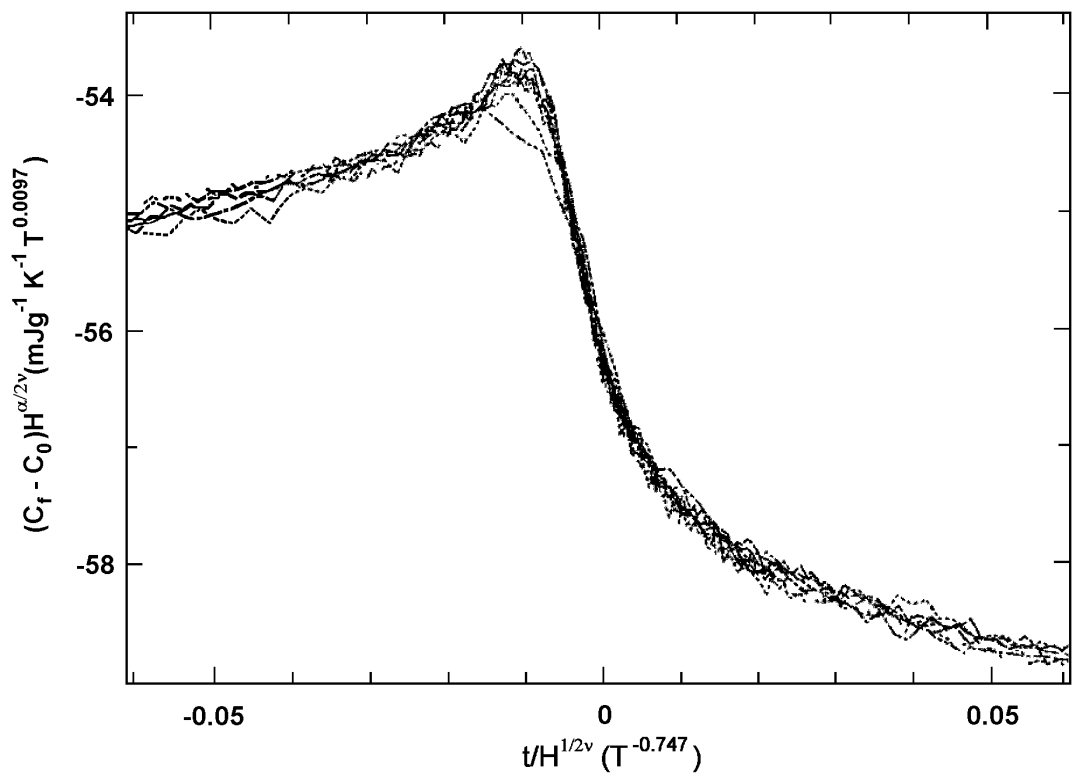
\begin{figure}
\caption{Experimental data for the specific-heat scaling function as reported in Ref. [4]. }
\label{fig4}
\end{figure}

\end{document}